\documentclass[bm,aps,
amsfonts,amssymb,
preprint,
nofootinbib]{revtex4}

\usepackage[dvipdfmx]{graphicx}
\usepackage{natbib}
\usepackage{amsmath}

\usepackage{color}

\font\mybb=msbm10 at 12pt

\font\mybbsub=msbm10 at 8pt
\font\mybbsmall=msbm10 at 10pt

\font\myeu=eufm10 at 12pt

\font\myeusmall=eufm10 at 10pt

\def\bb#1{\hbox{\mybb#1}}

\def\bbsub#1{\hbox{\mybbsub#1}}
\def\bbsmall#1{\hbox{\mybbsmall#1}}

\def\frak#1{\hbox{\myeu#1}}

\def\fraksmall#1{\hbox{\myeusmall#1}}

\def\ZZ {\bb{Z}}

\def\ZZsmall {\bbsmall{Z}}

\def\QQ {\bb{Q}}

\def\CCsub {\bbsub{C}}
\def\PP {\bb{P}}

\def\FF {\bb{F}}

\def\g {\frak{g}}

\def\gsmall {\fraksmall{g}}

\newcommand\beqa{\begin{eqnarray}}
\newcommand\eeqa{\end{eqnarray}}
\newcommand\n{\nonumber\\}

\begin{document}

{~}

\title{
Error-correcting codes over the Mordell-Weil groups of extremal rational elliptic surfaces and the $E_8$ lattice
}
\author{
Shun'ya Mizoguchi\footnote[1]{E-mail:mizoguch@post.kek.jp}
and Takumi Oikawa\footnotemark[2]{
}
}


\affiliation{\footnotemark[1]Theory Center, 
Institute of Particle and Nuclear Studies,
KEK\\Tsukuba, Ibaraki, 305-0801, Japan 
}

\affiliation{\footnotemark[1]\footnotemark[2]SOKENDAI (The Graduate University for Advanced Studies)\\
Tsukuba, Ibaraki, 305-0801, Japan 
}


\begin{abstract} 
We construct the $E_8$ lattice from classical error-correcting codes 
over the Mordell-Weil groups of rational elliptic surfaces that have
a singularity lattice of rank $8$ (maximal) for all cases of Oguiso-Shioda's 
classification.
By the structure theorem of the Mordell-Weil lattice of rational elliptic surfaces, 
if the rank of the singularity lattice is maximal,
 then the Mordell-Weil group is a cyclic group or a direct sum of them. 
 The singularity lattices are glued together by a code over their natural ring 
 to form the $E_8$ lattice.
Such constructions of the $E_8$ lattice from  codes can be seen as 
 a Lie algebraic extension and further generalization of 
 known code lattice constructions such as Construction A and Construction A${}_{\CCsub}$.
\end{abstract}

\preprint{KEK-TH-2816}
\date{March 8, 2026}

\maketitle

\section{Introduction}
In recent years, there has been a resurgence of interest in the relationship 
\cite{MacWilliamsSloane,ConwaySloan,Ebeling} 
between error-correcting codes and lattices, CFT, and 
string theory \cite{Dymarsky2020,
Dymarsky:2020bps,Dymarsky:2020pzc,Dymarsky2021,
Buican:2021uyp,Yahagi2022,Furuta:2022ykh,Henriksson2023,
Angelinos:2022umf,Henriksson:2022dml,Dymarsky:2022kwb,
Kawabata2023a,Furuta2023,Alam2023,Kawabata2023b,
Kawabata2024a,Buican:2023bzl,Aharony:2023zit,Barbar:2023ncl,Buican:2023ehi,
Ando:2024gcf,MizoguchiOikawa,Kawabata:2025hfd,Angelinos:2025mjj,
Ando:2025hwb,Dymarsky:2025agh,Keller:2025elq,
MizoguchiOikawa2}.
In this paper, we discuss the relationship between error-correcting codes 
that construct the $E_8$ lattice and the Mordell-Weil group of 
extremal rational elliptic surfaces.

In general, the construction of a lattice from an error-correcting code 
over a finite field, or ring such as a cyclic group (which has a natural 
ring structure),
can be viewed as a module homomorphism 
from the finite ring of the code 
to the quotient module of a lattice and its sublattice.
From this perspective,
to construct the $E_8$ lattice from a code, 
we may consider a sublattice of the $E_8$ lattice
whose quotient module is finite, i.e., a sublattice of rank $8$.
Such a setting can be realized exactly as a rational elliptic surface 
whose singularity lattice is maximal of rank $8$;
the quotient is the Mordell-Weil group.
The Mordell-Weil group of rational elliptic surfaces has been 
studied in detail by Oguiso-Shioda \cite{OguisoShioda}.

In this paper, we show what error correcting codes correspond to 
all cases of Oguiso-Shioda's classification.
After all, an error-correcting code specifies 
how the indices of the theta functions of orthogonal lattices 
shift relative to one another. 
The idea of constructing a modular invariant by adding all products 
of theta functions with indices shifted in this way 
was vigorously discussed 
in the context of string compactifications as the spectral flow \cite{spectralflow}
or the beta method in the Gepner model \cite{Gepner}.

This paper is organized as follows: 
In Section 2, we briefly review the Mordell-Weil lattices of rational elliptic surfaces.
In Section 3, we explain the concrete construction of the $E_8$ lattice from codes 
over the Mordell-Weil group, or its single direct-sum component, of extremal rational elliptic surfaces.
We examine three cases in turn, depending on whether the singularity lattice 
$T$ contains $D_{2N}=SO(4N)$ $(N=2,3,\ldots)$
in the direct-sum components, 
and if not, whether the map from the ring of the code to the quotient modules of the relevant lattices 
are all isomorphic or only homomorphic.
Finally, Section 4 summarizes the conclusions.

\section{Mordell-Weil lattices of rational elliptic surfaces
}
The Mordell-Weil group $E(K)$ of a rational elliptic surface 
(RES) is defined as the abelian group
of rational sections of rational elliptic surface, 
where $K$ is the field of rational functions of the coordinate $z$ 
of the base $\PP^1$ of rational elliptic surface.
The addition of two sections is defined by adding two points 
on an elliptic curve fiberwise.
According to \cite{OguisoShioda}, 
the Mordell-Weil (MW) group $E(K)$ of a rational elliptic surface 
is, under a suitable inner product, a lattice with the following structure:
\beqa
E(K)&\simeq&L^*\otimes (T'/T),
\eeqa
where $T$ is a singularity lattice embedded in the $E_8$ root lattice 
$\Lambda^{E_8}_R$,
$L$ is the orthogonal lattice $T^\perp$ of $T$ with respect to the embedding into $\Lambda^{E_8}_R$, and
\beqa
T'&=&T\otimes \QQ \cap \Lambda^{E_8}_R.
\eeqa 
If the singularity lattice has maximal rank $8$, then $L^*=1$ and the Mordell-Weil group $E(K)$ is finite.
They are listed in the table \ref{MW}.
\setlength{\tabcolsep}{10pt}
\renewcommand{\arraystretch}{0.6}
\setlength{\arraycolsep}{3pt}
\begin{table}[h]
\caption{ \label{MW}}
\centering
\small
\begin{tabular}{|c|c|c|}
\hline
No. &$T$ &$E(K)$\\
\hline
$63$ &$A_8$ &$\ZZ_3$\\
$64$ &$D_8$ &$\ZZ_2$\\
$65$ &$E_7\oplus A_1$ &$\ZZ_2$\\
$66$ &$A_5\oplus A_2 \oplus A_1$ &$\ZZ_6$\\
$67$ &$A_4\oplus A_4$ &$\ZZ_5$\\
$68$ &$A_2\oplus A_2\oplus A_2\oplus A_2$ &$\ZZ_3 \oplus \ZZ_3$\\
$69$ &$E_6\oplus A_2$ &$\ZZ_3$\\
$70$ &$A_7\oplus A_1$ &$\ZZ_4$\\
$71$ &$D_6\oplus A_1\oplus A_1$ &$\ZZ_2 \oplus \ZZ_2$\\
$72$ &$D_5\oplus A_3$ &$\ZZ_4$\\
$73$ &$D_4\oplus D_4$ &$\ZZ_2 \oplus \ZZ_2$\\
$74$ &$A_3\oplus A_1\oplus A_3\oplus A_1$ &$\ZZ_4 \oplus \ZZ_2$\\
\hline
\end{tabular}
\end{table}

As can be seen from this table,  $E(K)$ is either a cyclic group
or a direct sum of two cyclic groups in all these cases.
We will show that the direct-sum components of the singularity lattice 
are ``glued'' \cite{ConwaySloan} 
with a code over the Mordell-Weil group, or 
one over the cyclic group of one of its direct-sum components 
when it is a direct sum of cyclic groups, to form the 
$E_8$ lattice \footnote{
In No. 74, where $E(K)$ is a direct sum of different cyclic groups, 
we consider a code over $\ZZsmall_4$.
}.
In this paper we will find such glue codes for all cases in the table.

Looking at this procedure in reverse, we can see that 
it is a generalization of the construction of code lattices, 
such as Construction A${}_{\CCsub}$ which constructs the $E_8$ lattice from the tetracode.
Indeed, the construction of the $E_8$ lattice by the tetracode corresponds 
to No.~68 in the list of the Mordell-Weil group by Oguiso-Shioda \footnote{The construction of 
the $E_8$ lattice using the extended Hamming code does not correspond to an 
extremal rational elliptic surface. 
This is because the eight $I_2$ Kodaira fibers required for this construction require 
16 singular fibers, exceeding the total number of singular fibers $(=12)$ for rational elliptic surfaces.}.
Construction A${}_{\CCsub}$ can be said to be the operation of gluing together 
$SU(3)$ root lattices using a given code over $\FF_3$ as the glue code.
In \cite{MizoguchiOikawa}, we generalized this by taking a code on $\ZZ_k$ (or $\FF_k$) 
as a glue code, and gluing together the root lattices of a Lie algebra $\frak{g}$ 
whose dual quotient \cite{ConwaySloan} is $k$. 
We called this lattice construction ``Construction A${}_{\gsmall}$''.
In this paper, we further generalize this and make the following two extensions 
to construct the $E_8$ lattice from codes over the Mordell-Weil groups:
\begin{itemize}
\item[(i)]{\em Gluing root lattices of different Lie algebras with the same dual quotient}\\
In Construction A${}_{\gsmall}$ \cite{MizoguchiOikawa}, for a given code over 
$\ZZ_k$ (or $\FF_k$), we glued together the root lattices of a single Lie algebra $\frak{g}$
whose dual quotient \cite{ConwaySloan} is $k$, but it is  possible 
to glue together root lattices of different Lie algebras that have the same dual quotient.
In this way, we can realize glueings using glue codes 
that correspond to the orthogonal decomposition of $E_8$ 
familiar from the Higgs mechanism of the gauge groups of 
grand unified theories such as $SO(10)\times SU(4)$ and $E_6\times SU(3)$.

\item[(ii)]{\em Extension to the case where the map from the ring of codes to 
the quotient module of the weight and root lattices is not an isomorphism
but only a homomorphism}\\
In Construction A${}_{\gsmall}$, we assumed that the finite ring over which the code
is defined and the quotient module of the lattices are isomorphic, i.e., 
one-to-one, onto. Indeed, this was the case for the construction of 
code lattices using integers over cyclotomic fields \cite{MizoguchiOikawa}.  
From the perspective of constructing a lattice, however, we can relax this to a 
homomorphism with a non-trivial kernel or a cokernel.
By defining a lattice in this way, we can realize the $E_8$ lattice 
from codes for all classified extremal rational elliptic surfaces in the table.

\end{itemize}

In the following section, we present a concrete construction of 
the $E_8$ lattice from classical codes over the Mordell-Weil 
groups of extremal rational elliptic surfaces.
Our strategy is as follows:
We first consider a sum over the index shifts of the product 
of theta functions corresponding to the singularity lattice for each 
Oguiso-Shioda classification, determine the index shifts that 
make it modular invariant, and then read the glue code from it.
The uniqueness of the modular invariant theta \footnote{if 
supplemented by $\frac1{\eta(\tau)^8}$, as in the heterotic string 
partition function.} 
in eight Euclidean dimensions shows that it is the $E_8$ lattice.
In the case of binary codes, the modular invariance of the lattice theta 
obtained from Construction A corresponds to the doubly-even self-duality 
of the code, while in the individual cases below, the discussion will 
make clear what properties of the code imply the modular invariance 
of the lattice theta.

\vskip 5mm
\noindent
\underline{\em Remarks about Notation
}\\
In general, we define the theta function associated with the lattice $\Lambda$ by 
\beqa
\Theta_{\vec{\lambda},k}^{\Lambda}(\tau,\vec{z})&=&
\sum_{\vec{x}\in\Lambda}
q^{\frac k2 (\vec{x}+\frac{\vec{\lambda}}k)^2}
e^{2\pi i k \vec{z} (\vec{x}+\frac{\vec{\lambda}}k)}.
\eeqa
Let $\Lambda^{\gsmall}_R$ be the root lattice of the Lie algebra $\g$,
let $\Lambda^{\gsmall}_W$ be the weight lattice, and let $C_{\gsmall}$ be the Cartan matrix.
Then the formula for modular $S$-transformation is
\beqa
\Theta^{\Lambda^{\gsmall}_R}_{\vec{\lambda},k}(\tau,\vec{z})&=&\frac1{(-ik\tau)^{\frac r2}}
\frac{e^{-\frac{k\pi i}\tau
\vec{z}^2
}}{\det\sqrt{C_{\gsmall}}}
\sum_{\mu\in\Lambda^{\gsmall}_W/k\Lambda^{\gsmall}_R}
e^{2\pi i \frac{\vec{\lambda}\cdot\vec{\mu}}k}
\Theta^{\Lambda^{\gsmall}_R}_{\vec{\lambda},k}\left(-\frac1\tau,\frac{\vec{z}}\tau\right).
\eeqa
If $\vec{z}=0$, 
\beqa
\Theta^{\Lambda^{\gsmall}_R}_{\vec{\lambda},k}(\tau)&=&
\frac1{(-ik\tau)^{\frac r2}}
\frac
1{\sqrt{\det C_{\gsmall}}}
\sum_{\mu\in\Lambda^{\Lambda^{\gsmall}_R}_W/k\Lambda^{\gsmall}_R}
e^{2\pi i \frac{\vec{\lambda}\cdot\vec{\mu}}k}
\Theta^{\gsmall}_{\vec{\lambda},k}\left(-\frac1\tau\right).
\eeqa
In this paper, the level $k$ of the theta function is always set to $1$ and is 
omitted from the subscript of $\Theta^{\Lambda^{\gsmall}_R}$.

\vskip 5mm
\noindent
\underline{\em Remarks about Terminology
}\\
The codes we deal with in this paper are ones over $E(K)$ in table 1, 
or over one of its direct-sum components if it is a direct sum.
All of these codewords are composed of symbols whose values 
belong to some cyclic group, which has a natural ring structure 
induced from the ring of integers. 
We call this ring a {\em code ring}, to distinguish it from the {\em quotient module} 
of the weight lattice and the root lattice, 
which is also isomorphic to a cyclic group or a direct sum of them.

\section{Construction of the $E_8$ lattice 
from codes over the Mordell-Weil groups of extremal rational elliptic surfaces
}

As mentioned in the introduction, in this section we find codes that 
construct the $E_8$ lattice in the following three cases:
\\
Case 1: The case when the singularity lattice $T$ does not contain  
$D_{2N}=SO(4N)$ $(N=2,3,\ldots)$ in its direct-sum components, 
and the maps from the ring of the code to the quotient modules of the lattices are all isomorphic.\\
Case 2: The case when the singularity lattice $T$ does not contain  
$D_{2N}=SO(4N)$ $(N=2,3,\ldots)$ in its direct-sum components, 
and some of the maps from the ring of the code to the quotient modules of the lattices 
are only homomorphic.\\
Case 3: The case when the singularity lattice $T$ contains 
$D_{2N}=SO(4N)$ $(N=2,3,\ldots)$ in its direct-sum components. 
\\

\subsection*{Case 1
 }
We first consider the case when the singularity lattice $T$ does not contain  
$D_{2N}=SO(4N)$ $(N=2,3,\ldots)$ in its direct-sum components, 
and the maps from the code ring 
to the quotient module $\Lambda^{\gsmall}_W/\Lambda^{\gsmall}_R$ are all isomorphic.
Nos. 65, 67, 68, 69  and 72 belong to this class \footnote{
In this section, we will use notations such as $SU(N)$ and $SO(2N)$ for the Lie algebra $\g$ instead of $A_{N-1}$ and $D_{N}$.
This is to improve visibility as a superscript of the theta functions, and to address a wider audience familiar with particle physics, 
GUT, string theory, etc.
Also, Lie groups and Lie algebras are confused 
when the difference is clear from the context, as is often the case in the literature in these fields.
}.
Nos.67 and 68 are cases where the $E_8$ lattice is decomposed into sublattices 
of the same Lie algebra and can be reconstructed using Construction A$_{\gsmall}$ discussed 
in \cite{MizoguchiOikawa}. Nos. 65, 69 and 72 are the cases where the 
$E_8$ lattice is decomposed into those of different Lie algebras.

\subsection*{No.67 $SU(5)\times SU(5)$}
In this case, we consider
\beqa
\sum_{c=0}^4
\Theta_{\vec{c\omega_1}}^{\Lambda^{SU(5)}_R}(\tau)
\Theta_{\vec{c\omega_2}}^{\Lambda^{SU(5)}_R}(\tau).
\eeqa
This is the theta of the lattice constructed by Construction A$_{\gsmall}$
from the code over $\FF_5$ generated by 
\beqa
\left(
\begin{array}{cc}
1&2
\end{array}
\right).
\eeqa
The square of the row vector of this generator matrix is computed, by the usual Euclidean 
inner product, as
\beqa
1^2+2^2&=&5~\equiv~0~~~\mbox{mod $5$},
\eeqa
showing the the code is self-dual.
The modular $S$ transformation yields
\beqa
\sum_{c=0}^4
\Theta_{\vec{c\omega_1}}^{\Lambda^{SU(5)}_R}(\tau)
\Theta_{\vec{c\omega_2}}^{\Lambda^{SU(5)}_R}(\tau)&=&
\frac1{(-i\tau)^{\frac{4+4}2}}
\frac
1{\sqrt{5\cdot 5}}
\sum_{c=0}^4
\sum_{\vec{\mu}_1\in\Lambda^{SU(5)}_W/\Lambda^{SU(5)}_R}
\sum_{\vec{\mu}_2\in\Lambda^{SU(5)}_W/\Lambda^{SU(5)}_R}\n
&&\cdot
e^{2\pi i c(\vec{\omega_1}\cdot\vec{\mu_1}+
\vec{\omega_2}\cdot\vec{\mu_2})}
\Theta_{\vec{\mu_1}}^{\Lambda^{SU(5)}_R}\left(-\frac1\tau\right)
\Theta_{\vec{\mu_2}}^{\Lambda^{SU(5)}_R}\left(-\frac1\tau\right).
\label{SU(5)xSU(5)Theta}
\eeqa
Here
\beqa
\frac15
\sum_{c=0}^4
e^{2\pi i c(\vec{\omega}_1\cdot\vec{\mu}_1+
\vec{\omega}_2\cdot\vec{\mu}_2)}
\label{SU(5)phasesum}
\eeqa
is $\neq0$ only if
\beqa
\vec{\omega}_1\cdot\vec{\mu}_1+
\vec{\omega}_2\cdot\vec{\mu}_2
&\equiv&0~~~\mbox{mod $\ZZ$},
\eeqa
and then (\ref{SU(5)phasesum}) $=1$.
Since $\vec{\omega}_j\equiv j \vec{\omega}_1$ mod 
$\Lambda_R^{SU(N)}$ in $SU(N)$,  only such $(\vec{\mu}_1,\vec{\mu}_2)$ are
\beqa
(\vec{\mu_1},\vec{\mu_2})&=&c'(\vec{\omega}_1,\vec{\omega}_2)~~~
c'=0,1,2,3,4.
\eeqa
Thus
\beqa
\sum_{c=0}^4
\Theta_{c\vec{\omega}_1}^{\Lambda^{SU(5)}_R}(\tau)
\Theta_{c\vec{\omega}_2}^{\Lambda^{SU(5)}_R}(\tau)&=&
\sum_{c'=0}^4
\frac1{(-i\tau)^{4}}
\Theta_{c\vec{\omega}_1}^{\Lambda^{SU(5)}_R}\left(-\frac1\tau\right)
\Theta_{c\vec{\omega}_2}^{\Lambda^{SU(5)}_R}\left(-\frac1\tau\right),
\eeqa
which, together with $\frac1{\eta(\tau)^8}$, is modular $S$ invariant.
This is consistent with the fact that the code is self-dual.  
Also, in the case of $SU(5)$, $(\vec
\omega_1)^2$ is $\frac45$ whose  numerator is even,  so
 as a column vector, even if the square of  the code is not  a``doubly multiple of $5$'', 
 it is modular $T$ invariant as long as the square is simply a multiple of $5$.
(In Construction A for binary codes,
the $(\vec
\omega_1)^2$ of $SU(2)$ is $=\frac12$, so doubly even is a condition for modular $T$ invariance.)
Therefore, due to the uniqueness of eight-dimensional modular invariant lattices (= even self-dual lattices), 
this is the $E_8$ lattice.

\subsection*{No.68 $SU(3)\times SU(3)\times SU(3)\times SU(3)$}
As another example of decomposition into a direct product of 
lattices of the same Lie algebra, 
let us consider $E_8\supset SU(3)\times SU(3)\times SU(3)\times SU(3)$:
\beqa
\sum_{c_1=0}^2\sum_{c_2=0}^2
\Theta_{c_1\vec{\omega}_1}^{\Lambda^{SU(3)}_R}
\Theta_{c_2\vec{\omega}_1}^{\Lambda^{SU(3)}_R}
\Theta_{(c_1+c_2)\vec{\omega}_1}^{\Lambda^{SU(3)}_R}
\Theta_{(c_1-c_2)\vec{\omega}_1}^{\Lambda^{SU(3)}_R}(\tau).
\eeqa
By performing a modular $S$ transformation, we find
\beqa
&&
\sum_{c_1=0}^2\sum_{c_2=0}^2
\Theta_{c_1\vec{\omega}_1}^{\Lambda^{SU(3)}_R}
\Theta_{c_2\vec{\omega}_1}^{\Lambda^{SU(3)}_R}
\Theta_{(c_1+c_2)\vec{\omega}_1}^{\Lambda^{SU(3)}_R}
\Theta_{(c_1-c_2)\vec{\omega}_1}^{\Lambda^{SU(3)}_R}(\tau)\n
&=&
\frac1{(-i\tau)^{\frac{2+2+2+2}2}}
\frac
1{\sqrt{3}^4}
\sum_{c_1=0}^2\sum_{c_2=0}^2
\sum_{\vec{\mu}_1\in\Lambda^{SU(3)}_W/\Lambda^{SU(3)}_R}
\sum_{\vec{\mu}_2\in\Lambda^{SU(3)}_W/\Lambda^{SU(3)}_R}
\sum_{\vec{\mu}_3\in\Lambda^{SU(3)}_W/\Lambda^{SU(3)}_R}
\sum_{\vec{\mu}_4\in\Lambda^{SU(3)}_W/\Lambda^{SU(3)}_R}
\n
&&\cdot
e^{2\pi i (c_1\vec{\omega}_1\cdot\vec{\mu}_1+
c_2\vec{\omega}_1\cdot\vec{\mu}_2+
(c_1+c_2)\vec{\omega}_1\cdot\vec{\mu}_3+
(c_1-c_2)\vec{\omega}_1\cdot\vec{\mu}_4)
}
\Theta_{\vec{\mu}_1}^{\Lambda^{SU(3)}_R}
\Theta_{\vec{\mu}_2}^{\Lambda^{SU(3)}_R}
\Theta_{\vec{\mu}_3}^{\Lambda^{SU(3)}_R}
\Theta_{\vec{\mu}_4}^{\Lambda^{SU(3)}_R}\left(-\frac1\tau\right).
\label{SU(3)^4Theta}
\eeqa

\beqa
&&\frac
1{\sqrt{3}^4}
\sum_{c_1=0}^2\sum_{c_2=0}^2
e^{2\pi i (c_1\vec{\omega}_1\cdot\vec{\mu}_1+
c_2\vec{\omega}_1\cdot\vec{\mu}_2+
(c_1+c_2)\vec{\omega}_1\cdot\vec{\mu}_3+
(c_1-c_2)\vec{\omega}_1\cdot\vec{\mu}_4)
}
\n
&=&
\frac13
\sum_{c_1=0}^2
e^{2\pi i c_1(\vec{\omega}_1\cdot\vec{\mu}_1+
\vec{\omega}_1\cdot\vec{\mu}_3+
\vec{\omega}_1\cdot\vec{\mu}_4)
}
\frac13
\sum_{c_2=0}^2
e^{2\pi i c_2(
\vec{\omega}_1\cdot\vec{\mu}_2+
\vec{\omega}_1\cdot\vec{\mu}_3-
\vec{\omega}_1\cdot\vec{\mu}_4)
}
\label{deltaSU(3)^4}
\eeqa
with
\beqa
\vec{\mu}_1=m'_1\vec{\omega}_1,
~~~
\vec{\mu}_2=m'_2\vec{\omega}_1,
~~~
\vec{\mu}_3=m'_3\vec{\omega}_1,
~~~
\vec{\mu}_4=m'_4\vec{\omega}_1
~~~
\eeqa
is $\neq 0$ iff, since $\vec{\omega}_1^2=\frac23$, 
\beqa
\frac23 c_1(m'_1+m'_3+m'_4)&\equiv&0~~~\mbox{mod $\ZZ$},\n
\frac23 c_2(m'_2+m'_3-m'_4)&\equiv&0~~~\mbox{mod $\ZZ$},
\eeqa
which means that 
\beqa
c_1(m'_1+m'_3+m'_4)&\equiv&0~~~\mbox{mod $3$},\n
c_2(m'_2+m'_3-m'_4)&\equiv&0~~~\mbox{mod $3$},
\eeqa
are satisfied simultaneously,
and then $ (\ref{deltaSU(3)^4})=1$.
These conditions can be written as
\beqa
\left(
\begin{array}{cccc}
1~&~0~&~1&~~1\\
0~&~1~&~1&-1
\end{array}
\right)
\left(
\begin{array}{c}
m'_1\\
m'_2\\
m'_3\\
m'_4
\end{array}
\right)
&\equiv
&
\left(
\begin{array}{c}
0\\
0
\end{array}
\right)~~~\mbox{mod $3$}.
\eeqa
The matrix on the left hand side is the generator matrix of the tetracode, which is self-dual,
so $(m'_1,m'_2,m'_3,m'_4)$ is given by a $\ZZ_3$-coefficient linear combination 
of the two rows $(1,0,1,1)$ and $(0,1,1,-1)$:
\beqa
\left(
\begin{array}{cccc}
m'_1&
m'_2&
m'_3&
m'_4
\end{array}
\right)
&=&
\left(
\begin{array}{cc}
c'_1&
c'_2
\end{array}
\right)
\left(
\begin{array}{cccc}
1~&~0~&~1&~~1\\
0~&~1~&~1&-1
\end{array}
\right)
~~~~~~c'_1,c'_2\in\ZZ_3 .
\eeqa
Therefore
\beqa
&&
\sum_{c_1=0}^2\sum_{c_2=0}^2
\Theta_{c_1\vec{\omega}_1}^{\Lambda^{SU(3)}_R}
\Theta_{c_2\vec{\omega}_1}^{\Lambda^{SU(3)}_R}
\Theta_{(c_1+c_2)\vec{\omega}_1}^{\Lambda^{SU(3)}_R}
\Theta_{(c_1-c_2)\vec{\omega}_1}^{\Lambda^{SU(3)}_R}
(\tau)\n
&=&
\frac1{(-i\tau)^4}
\sum_{c'_1=0}^2\sum_{c'_2=0}^2
\Theta_{c'_1\vec{\omega}_1}^{\Lambda^{SU(3)}_R}
\Theta_{c'_2\vec{\omega}_1}^{\Lambda^{SU(3)}_R}
\Theta_{(c'_1+c'_2)\vec{\omega}_1}^{\Lambda^{SU(3)}_R}
\Theta_{(c'_1-c'_2)\vec{\omega}_1}^{\Lambda^{SU(3)}_R}
\left(-\frac1\tau\right).
\eeqa
In this case, this (with $\frac1{\eta(\tau)^8}$) is also modular $T$ invariant because 
$(\vec\omega_1)^2=\frac23$  in $SU(3)$, whose numerator is even,  
and when each column of codes is viewed as a vector, its square is $3$.   

The code in this case is generated by
\beqa
\left(
\begin{array}{cccc}
1~&~0~&~1&~~1\\
0~&~1~&~1&-1
\end{array}
\right),
\eeqa
that is, this is the tetracode.

\subsection*{No.69 $SU(3)\times E_6$}
Let us next  consider a code over $\FF_3$($\ZZ_3$) 
generated by the generator matrix 
\beqa
\left(
\begin{array}{cc}
1&1
\end{array}
\right),
\label{(11)}
\eeqa
and construct a lattice by Construction A$_{\gsmall}$
using the $SU(3)$ lattice for the first symbol, and 
the $E_6$ lattice for the second symbol.
Since $\Lambda_W/\Lambda_R=\ZZ_3$ in both $SU(3)$ and $E_6$, 
this ``mixed'' Construction A$_{\gsmall}$ is possible.
The lattice theta is
\beqa
\sum_{c=0}^2
\Theta_{c\vec{\omega}^{SU(3)}_1}^{\Lambda^{SU(3)}_R}(\tau)
\Theta_{c\vec{\omega}^{E_6}_1}^{\Lambda^{E_6}_R}(\tau),
\eeqa
whose modular $S$ transformation reads
\beqa
\sum_{c=0}^2
\Theta_{c\vec{\omega}^{SU(3)}_1}^{\Lambda^{SU(3)}_R}(\tau)
\Theta_{c\vec{\omega}^{E_6}_1}^{\Lambda^{E_6}_R}(\tau)
&=&
\frac1{(-i\tau)^{\frac{2+6}2}}
\frac
1{\sqrt{3\cdot 3}}
\sum_{c=0}^2
\sum_{\vec{\mu}_1\in\Lambda^{SU(3)}_W/\Lambda^{SU(3)}_R}
\sum_{\vec{\mu}_2\in\Lambda^{E_6}_W/\Lambda^{E_6}_R}\n
&&\cdot
e^{2\pi i c(\vec{\omega}_1^{SU(3)}\cdot\vec{\mu}_1+
\vec{\omega}_1^{E_6}\cdot\vec{\mu}_2)}
\Theta_{\vec{\mu}_1}^{\Lambda^{SU(3)}_R}\left(-\frac1\tau\right)
\Theta_{\vec{\mu}_2}^{\Lambda^{E_6}_R}\left(-\frac1\tau\right).
\label{SU(3)xE6Theta}
\eeqa
Here the factor
\beqa
\frac13
\sum_{c=0}^2
e^{2\pi i c(\vec{\omega}_1^{SU(3)}\cdot\vec{\mu}_1+
\vec{\omega}_1^{E_6}\cdot\vec{\mu}_2)}
\label{deltaSU(3)E6}
\eeqa
is $\neq 0$, $=1$ iff
\beqa
\vec{\mu}_1=\vec{\omega}_1^{SU(3)},~~~
\vec{\mu}_2=\vec{\omega}_1^{E_6}
\label{omega^2SU(3)E6}
\eeqa
as
\beqa
(\vec{\omega}_1^{SU(3)})^2=\frac23,~~~
(\vec{\omega}_1^{E_6})^2=\frac43.
\eeqa
Therefore, we find
\beqa
\sum_{c=0}^2
\Theta_{c\vec{\omega}^{SU(3)}_1}^{\Lambda^{SU(3)}_R}(\tau)
\Theta_{c\vec{\omega}^{E_6}_1}^{\Lambda^{E_6}_R}(\tau)
&=&
\sum_{c'=0}^4
\frac1{(-i\tau)^{4}}
\Theta_{c\vec{\omega}^{SU(3)}_1}^{\Lambda^{SU(3)}_R}\left(-\frac1\tau\right)
\Theta_{c\vec{\omega}^{E_6}_1}^{\Lambda^{E_6}_R}\left(-\frac1\tau\right),
\label{SU(3)xE6Theta}
\eeqa
\noindent
which is modular invariant if it is multiplied by $\frac1{\eta(\tau)^8}$.

In the normal Euclidean inner product, (\ref{(11)}) does not have its length squared $0$ mod $3$.
This is because the squared lengths of the generators of 
$\Lambda_W/\Lambda_R$ of the two Lie algebras 
used in Construction A$_{\gsmall}$ are different.
In order to maintain the relationship between the square of the length 
of the row vector of the generator matrix and modular invariance, 
we simply define, due to (\ref{deltaSU(3)E6}),
 (\ref{omega^2SU(3)E6}),  the inner product as  
\beqa
 \left(
\begin{array}{cc}
c_1&c_2
\end{array}
\right)
 \left(
\begin{array}{cc}
1&\\&2
\end{array}
\right)
 \left(
\begin{array}{c}
c_1\\c_2
\end{array}
\right)
 \eeqa
 for a codeword 
  \beqa
 \left(
\begin{array}{cc}
c_1&c_2
\end{array}
\right)
 \eeqa
 ($c_1, c_2\in \FF_3$).
 In this case, $c_1=c_2=1$, so the value of this inner product is $0$ mod $3$.
The $2$ in the inner product matrix reflects the fact that 
$(\vec{\omega}_1^{E_6})^2$ is $2$ times $(\vec{\omega}_1^{SU(3)})^2$.
Again, (\ref{SU(3)xE6Theta}) is modular $T$ invariant without needing to be a``doubly multiple of three'',
unlike binary codes.

\subsection*{No.65 $SU(2)\times E_7$}
Since $\Lambda_W/\Lambda_R$ of both of Lie algebras are also equal $=\ZZ_2$, 
the $E_8$ lattice can be constructed by Construction A$_{\gsmall}$ using root lattices of these Lie algebras, 
just like $SU(3)\times E_6$.
The lattice theta we consider in this case is
\beqa
\sum_{c=0}^1
\Theta_{c\vec{\omega}^{SU(2)}_1}^{\Lambda^{SU(2)}_R}(\tau)
\Theta_{c\vec{\omega}^{E_7}_1}^{\Lambda^{E_7}_R}(\tau),
\label{SU(2)xE7Theta}
\eeqa
where $\vec{\omega}^{E_7}_1$ is the fundamental weight corresponding 
to the node connected to the node removed from the extended Dynkin diagram of $E_8$.
Its modular $S$ transform reads 
\beqa
(\ref{SU(2)xE7Theta})
&=&
\frac1{(-i\tau)^{\frac{1+7}2}}
\frac
1{\sqrt{2\cdot 2}}
\sum_{c=0}^1
\sum_{\vec{\mu}_1\in\Lambda^{SU(2)}_W/\Lambda^{SU(2)}_R}
\sum_{\vec{\mu}_2\in\Lambda^{E_7}_W/\Lambda^{E_7}_R}\n
&&\cdot
e^{2\pi i c(\vec{\omega}_1^{SU(2)}\cdot\vec{\mu}_1+
\vec{\omega}_1^{E_7}\cdot\vec{\mu}_2)}
\Theta_{\vec{\mu}_1}^{\Lambda^{SU(2)}_R}\left(-\frac1\tau\right)
\Theta_{\vec{\mu}_2}^{\Lambda^{E_7}_R}\left(-\frac1\tau\right).
\eeqa
Since
\beqa
(\vec{\omega}_1^{SU(2)})^2=\frac12,~~~
(\vec{\omega}_1^{E_7})^2=\frac32,
\eeqa
by setting 
$\vec{\mu}_1=c_1\,\vec{\omega}_1^{SU(2)}$,
$\vec{\mu}_2=c_2\,\vec{\omega}_1^{E_7}$,
we have
\beqa
\frac12
\sum_{c=0}^1
e^{2\pi i c(\vec{\omega}_1^{SU(2)}\cdot\vec{\mu}_1+
\vec{\omega}_1^{E_7}\cdot\vec{\mu}_2)}
&=&
\frac12
\sum_{c=0}^1
e^{2\pi i c(\frac23 c_1+ \frac43 c_2)},
\eeqa
which is $=1\neq 0$ if
\beqa
(c_1, c_2)&=&c'(1,1)~~~(c'=0,1).
\eeqa
Therefore, (\ref{SU(2)xE7Theta}) is modular $S$ invariant if we divide it by $\eta(\tau)^8$.
The generator matrix is 
\beqa
\left(
\begin{array}{cc}
1~&1
\end{array}
\right),
\eeqa
whose length squared is
\beqa
 \left(
\begin{array}{cc}
1&1
\end{array}
\right)
 \left(
\begin{array}{cc}
1&\\&3
\end{array}
\right)
 \left(
\begin{array}{c}
1\\1
\end{array}
\right)
&=&4,
 \eeqa
 showing that the code is doubly even. Thus (\ref{SU(2)xE7Theta}) is also modular $T$ invariant.

\subsection*{No.72 $SU(4)\times SO(10)$}
Since each of these Lie algebra factors also has the same $\Lambda_W/\Lambda_R=\ZZ_4$,
we can similarly construct the $E_8$ lattice by Construction A$_{\gsmall}$ using root lattices of different Lie algebras.
The generator of $\Lambda_W/\Lambda_R$ of $SO(10)$ is one of the fundamental weights 
of the two spinner representations; here we take $\vec{\omega}_4^{SU(10)}$.
The modular $S$ transformation of the lattice theta is
\beqa
\sum_{c=0}^3
\Theta_{c\vec{\omega}^{SU(4)}_1}^{\Lambda^{SU(4)}_R}(\tau)
\Theta_{c\vec{\omega}^{SO(10)}_4}^{\Lambda^{SO(10)}_R}(\tau)
&=&
\frac1{(-i\tau)^{\frac{3+5}2}}
\frac
1{\sqrt{4\cdot 4}}
\sum_{c=0}^3
\sum_{\vec{\mu}_1\in\Lambda^{SU(4)}_W/\Lambda^{SU(4)}_R}
\sum_{\vec{\mu}_2\in\Lambda^{SO(10)}_W/\Lambda^{SO(10)}_R}\n
&&\cdot
e^{2\pi i c(\vec{\omega}_1^{SU(4)}\cdot\vec{\mu}_1+
\vec{\omega}_1^{SO(10)}\cdot\vec{\mu}_2)}
\Theta_{\vec{\mu}_1}^{\Lambda^{SU(4)}_R}\left(-\frac1\tau\right)
\Theta_{\vec{\mu}_2}^{\Lambda^{SO(10)}_R}\left(-\frac1\tau\right).
\label{SU(4)xSO(10)Theta}
\eeqa
The lengths squared of the fundamental weights are
\beqa
(\vec{\omega}_1^{SU(4)})^2=\frac34,~~~
(\vec{\omega}_1^{SO(10})^2=\frac54,
\eeqa
so if we set
$\vec{\mu}_1=c_1\,\vec{\omega}_1^{SU(4)}$,
$\vec{\mu}_2=c_2\,\vec{\omega}_1^{SO(10)}$,
we obtain
\beqa
\frac14
\sum_{c=0}^3
e^{2\pi i c(\vec{\omega}_1^{SU(4)}\cdot\vec{\mu}_1+
\vec{\omega}_1^{SO(10)}\cdot\vec{\mu}_2)}
&=&
\frac12
\sum_{c=0}^1
e^{2\pi i c(\frac34 c_1+ \frac54 c_2)}.
\eeqa
This is $=1\neq 0$ if 
\beqa
(c_1, c_2)&=&c'(1,1)~~~(c'=0,1,2,3),
\eeqa
so (\ref{SU(4)xSO(10)Theta}) is also modular $S$ invariant if divided by $\eta(\tau)^8$.
The generator matrix is again
\beqa
\left(
\begin{array}{cc}
1~&1
\end{array}
\right),
\eeqa
whose length squared is computed to be
\beqa
 \left(
\begin{array}{cc}
1&1
\end{array}
\right)
 \left(
\begin{array}{cc}
3&\\&5
\end{array}
\right)
 \left(
\begin{array}{c}
1\\1
\end{array}
\right)
&=&8,
 \eeqa
 which is a``doubly multiple of four''.
This ensures the modular $T$ invariance of (\ref{SU(4)xSO(10)Theta}).

\subsection*{Case 2
 }
We next consider the case when the singularity lattice $T$ does not contain  
$D_{2N}=SO(4N)$, but  
some of the maps from the code ring to the quotient modules of the lattices 
are only homomorphic. 
Cases that fall under this category are Nos. 63, 66, 70 and 74.

\subsection*{No.63 $SU(9)$}
First, as an example of a one-to-one non-surjective map, i.e., the case when the cokernel is not $0$, 
consider the case where the extremal RES has an $SU(9)$($=I_9$) fiber.
In this case the MW group is $\ZZ_3$, while $\Lambda^{SU(9)}_W/\Lambda^{SU(9)}_R=\ZZ_9$,
so they do not coincide. However, let $\vec{\omega}_1$ be the fundamental weight 
that generates $\Lambda^{SU(9)}_W/\Lambda^{SU(9)}_R$, then 
$\{0,3\vec{\omega}_1,6\vec{\omega}_1\}$ form a submodule $\ZZ_3$.
We will use the homomorphism 
\beqa
c\in\ZZ_3~~~\mapsto~~~3c\vec{\omega}_1~\in~\Lambda^{SU(9)}_W/\Lambda^{SU(9)}_R
\eeqa
from $\ZZ_3$ to $\Lambda^{SU(9)}_W/\Lambda^{SU(9)}_R$ to construct a code lattice.
The lattice theta is
\beqa
\sum_{c=0}^2
\Theta_{3c\,\vec{\omega}_1}^{\Lambda^{SU(9)}_R}(\tau).
\eeqa
If we perform the modular $S$ transformation, we find
\beqa
\sum_{c=0}^2
\Theta_{3c\,\vec{\omega}_1}^{\Lambda^{SU(9)}_R}(\tau)
&=&
\frac1{(-i\tau)^{\frac{8}2}}
\frac
1{\sqrt{9}}
\sum_{c=0}^2
\sum_{\vec{\mu}\in\Lambda^{SU(9)}_W/\Lambda^{SU(9)}_R}
e^{2\pi i \cdot 3c\vec{\omega_1}\cdot\vec{\mu}}
\Theta_{\vec{\mu}}^{\Lambda^{SU(9)}_R}\left(-\frac1\tau\right).
\label{SU(9)Theta}
\eeqa
Plugging 
$\vec{\mu}=m'\vec{\omega}_1$, $m'\in\ZZ_9$ into this, we have
\beqa
&=&
\frac1{(-i\tau)^4}
\frac
13
\sum_{c=0}^2
\sum_{
m'=0}^8
e^{2\pi i \cdot \frac{8}{3} c m'}
\Theta_{m'\vec{\omega}_1}^{\Lambda^{SU(9)}_R}\left(-\frac1\tau\right)
\eeqa
as $\vec{\omega}_1^2=\frac89$. 
Since 
$\frac
13
\sum_{c=0}^2
e^{2\pi i \cdot \frac{8}{3} c m'}
=1\neq0 $ 
if $m'=0,3,6$, this becomes
\beqa
&=&
\frac1{(-i\tau)^4}
\sum_{c'=0}^2
\Theta_{3c'\vec{\omega}_1}^{\Lambda^{SU(9)}_R}\left(-\frac1\tau\right),
\eeqa
which is modular $S$ invariant together with $\frac1{(-i\tau)^4}$.
In this case, the generator matrix of the code over $\ZZ_3$ is simply 
\beqa
(1),
\eeqa
while as a code over $\ZZ_9$, it is  
\beqa
(3),
\eeqa
which satisfies $3^2=9\equiv0$ mod $9$ so the code is  self-dual.
Again, the square of the length does not need to be a ``doubly multiple of $9$''
 for the theta to be modular $T$ invariant.  

 \subsection*{No.66 $SU(6)\times SU(3)\times SU(2)$}
The lattice theta in this case is
 \beqa
\sum_{c=0}^5
\Theta_{c\vec{\omega}^{SU(6)}_1}^{\Lambda^{SU(6)}_R}(\tau)
\Theta_{c\vec{\omega}^{SU(3)}_1}^{\Lambda^{SU(3)}_R}(\tau)
\Theta_{c\vec{\omega}^{SU(2)}_1}^{\Lambda^{SU(2)}_R}(\tau),
\eeqa
which is modular-$S$ transformed to
 \beqa
 &=&
\frac1{(-i\tau)^{\frac{5+2+1}2}}
\frac
1{\sqrt{6\cdot 3 \cdot 2}}
\sum_{c=0}^5
\sum_{\vec{\mu}_1\in\Lambda^{SU(6)}_W/\Lambda^{SU(6)}_R}
\sum_{\vec{\mu}_2\in\Lambda^{SU(3)}_W/\Lambda^{SU(3)}_R}
\sum_{\vec{\mu}_3\in\Lambda^{SU(2)}_W/\Lambda^{SU(2)}_R}\n
&&\cdot
e^{2\pi i c(\vec{\omega}_1^{SU(6)}\cdot\vec{\mu}_1+
\vec{\omega}_1^{SU(3)}\cdot\vec{\mu}_2+
\vec{\omega}_1^{SU(2)}\cdot\vec{\mu}_3)}
\Theta_{\vec{\mu}_1}^{\Lambda^{SU(6)}_R}\!\!\left(-\frac1\tau\right)
\Theta_{\vec{\mu}_2}^{\Lambda^{SU(3)}_R}\!\!\left(-\frac1\tau\right)
\Theta_{\vec{\mu}_3}^{\Lambda^{SU(2)}_R}\!\!\left(-\frac1\tau\right).
\label{SU(6)xSU(3)xSU(2)Theta}
\eeqa
Let
\beqa
\vec{\mu}_1=c_1\,\vec{\omega}^{SU(6)}_1,~~~
\vec{\mu}_2=c_2\,\vec{\omega}^{SU(3)}_1,~~~
\vec{\mu}_3=c_3\,\vec{\omega}^{SU(2)}_1
~~~
(c_1\in\ZZ_6,~c_2\in\ZZ_3,~c_2\in\ZZ_2),
\eeqa
then due to the fact that
\beqa
(\vec{\omega}^{SU(6)}_1)^2~=\frac56,
~~~
(\vec{\omega}^{SU(3)}_1)^2~=\frac32,
~~~
(\vec{\omega}^{SU(2)}_1)^2~=\frac12,
\eeqa
we obtain
 \beqa
&&\frac
1{\sqrt{6\cdot 3 \cdot 2}}
\sum_{c=0}^5
e^{2\pi i c(\vec{\omega}_1^{SU(6)}\cdot\vec{\mu}_1+
\vec{\omega}_1^{SU(3)}\cdot\vec{\mu}_2+
\vec{\omega}_1^{SU(2)}\cdot\vec{\mu}_3)}
\n
&=&\frac
16\sum_{c=0}^5
e^{2\pi i c(\frac56 c_1+\frac32 c_2+\frac21 c_3).
}
\eeqa
It can be shown that this is $=1\neq 0$ if
\beqa
(c_1, c_2, c_3)&=&c'(1,1,1)~~~(c'=0,1,2,3,4,5).
\eeqa
Therefore
\beqa
&&\sum_{c=0}^5
\Theta_{c\vec{\omega}^{SU(8)}_1}^{\Lambda^{SU(6)}_R}(\tau)
\Theta_{c\vec{\omega}^{SU(2)}_1}^{\Lambda^{SU(3)}_R}(\tau)
\Theta_{c\vec{\omega}^{SU(2)}_1}^{\Lambda^{SU(2)}_R}(\tau)
\n
&=&
\frac1{(-i\tau)^{4}}
\sum_{c'=0}^5
\Theta_{c'\vec{\omega}^{SU(6)}_1}^{\Lambda^{SU(6)}_R}\!\!\left(-\frac1\tau\right)
\Theta_{c'\vec{\omega}^{SU(3)}_1}^{\Lambda^{SU(3)}_R}\!\!\left(-\frac1\tau\right)
\Theta_{c'\vec{\omega}^{SU(2)}_1}^{\Lambda^{SU(2)}_R}\!\!\left(-\frac1\tau\right).
\eeqa
This is a lattice summation obtained by Construction A${}_{\gsmall}$
using a code over $\ZZ_6$ with the generator matrix
\beqa
\left(
\begin{array}{ccc}
1~&1~&1
\end{array}
\right),
\eeqa
where $\g=SU(6)$ for the first symbol, $\g=SU(3)$ for the second symbol, 
and $\g=SU(2)$ for the third symbol.
The inner product of the rows of the generator matrix is defined as
\beqa
 \left(
\begin{array}{ccc}
1~&1~&1
\end{array}
\right)
\left(
\begin{array}{ccc}
5&&\\
&4&\\
&&3
\end{array}
\right)
 \left(
\begin{array}{c}
1\\1\\1
\end{array}
\right)
&=&12~\equiv 0~~~\mbox{mod $6$},
 \eeqa
 showing that the code is self-dual, and a ``doubly'' multiple of $6$. 
In this case, for the first symbol,
$\ZZ_6\rightarrow\Lambda^{SU(6)}_W/\Lambda^{SU(6)}_R$
is an isomorphism,
while for the second and third symbols,
the homomorphisms $\ZZ_6\rightarrow\Lambda^{SU(3)}_W/\Lambda^{SU(3)}_R$ and
$\ZZ_6\rightarrow\Lambda^{SU(2)}_W/\Lambda^{SU(2)}_R$
are the cases with kernels.

\subsection*{No.70 $SU(8)\times SU(2)$}
The MW group in this case is $\ZZ_4$.
On the other hand, $\Lambda_W/\Lambda_R$ in $SU(8)$ is $\ZZ_8$
and $\Lambda_W/\Lambda_R$ in $SU(2)$ is $\ZZ_2$,
so the homomorphism $\ZZ_4\rightarrow\Lambda^{SU(8)}_W/\Lambda^{SU(8)}_R$
has a cokernel, and
the homomorphism $\ZZ_4\rightarrow\Lambda^{SU(2)}_W/\Lambda^{SU(2)}_R$
has a kernel. We consider the lattice theta
\beqa
\sum_{c=0}^3
\Theta_{2c\vec{\omega}^{SU(8)}_1}^{\Lambda^{SU(8)}_R}(\tau)
\Theta_{c\vec{\omega}^{SU(2)}_1}^{\Lambda^{SU(2)}_R}(\tau),
\eeqa
whose modular $S$ transformation is found to be
\beqa
\sum_{c=0}^3
\Theta_{2c\vec{\omega}^{SU(8)}_1}^{\Lambda^{SU(8)}_R}(\tau)
\Theta_{c\vec{\omega}^{SU(2)}_1}^{\Lambda^{SU(2)}_R}(\tau)
&=&
\frac1{(-i\tau)^{\frac{7+1}2}}
\frac
1{\sqrt{8\cdot 2}}
\sum_{c=0}^3
\sum_{\vec{\mu}_1\in\Lambda^{SU(8)}_W/\Lambda^{SU(8)}_R}
\sum_{\vec{\mu}_2\in\Lambda^{SU(2)}_W/\Lambda^{SU(2)}_R}\n
&&\cdot
e^{2\pi i (2c\,\vec{\omega}_1^{SU(8)}\cdot\vec{\mu}_1+
c\,\vec{\omega}_1^{SU(2)}\cdot\vec{\mu}_2)}
\Theta_{\vec{\mu}_1}^{\Lambda^{SU(8)}_R}\!\!\left(-\frac1\tau\right)
\Theta_{\vec{\mu}_2}^{\Lambda^{SU(2)}_R}\!\!\left(-\frac1\tau\right).
\label{SU(2)xSU(8)Theta}
\eeqa
Let 
\beqa
\vec{\mu}_1=c_1\,\vec{\omega}^{SU(8)}_1,~~~
\vec{\mu}_2=c_2\,\vec{\omega}^{SU(2)}_1
~~~
(c_1\in\ZZ_8,~c_2\in\ZZ_2),
\eeqa
then 
\beqa
(\vec{\omega}^{SU(8)}_1)^2~=\frac78,
~~~
(\vec{\omega}^{SU(2)}_1)^2~=\frac12
\eeqa
imply
\beqa
\frac
1{\sqrt{8\cdot 2}}
\sum_{c=0}^3
e^{2\pi i (2c\,\vec{\omega}_1^{SU(8)}\cdot\vec{\mu}_1+
c\,\vec{\omega}_1^{SU(2)}\cdot\vec{\mu}_2)}
&=&
\frac
14
\sum_{c=0}^3
e^{2\pi i c
(\frac74 c_1 +\frac12 c_2)}.
\eeqa
This is $=1\neq 0$ if 
\beqa
(c_1, c_2)&=&c'(2,1)~~~(c'=0,1,2,3),
\eeqa
so that
\beqa
\sum_{c=0}^3
\Theta_{2c\vec{\omega}^{SU(8)}_1}^{\Lambda^{SU(8)}_R}(\tau)
\Theta_{c\vec{\omega}^{SU(2)}_1}^{\Lambda^{SU(2)}_R}(\tau)
&=&
\frac1{(-i\tau)^{4}}
\sum_{c'=0}^3
\Theta_{2c'\vec{\omega}^{SU(8)}_1}^{\Lambda^{SU(8)}_R}\!\!\left(-\frac1\tau\right)
\Theta_{c'\vec{\omega}^{SU(2)}_1}^{\Lambda^{SU(2)}_R}\!\!\left(-\frac1\tau\right).
\eeqa
This is the theta of the lattice of a code over $\ZZ_4$ generated by the generator matrix
\beqa
\left(
\begin{array}{cc}
2~&1
\end{array}
\right)
\eeqa
constructed by Construction A${}_{\gsmall}$, where 
$\g=SU(8)$ for the first symbol, and $\g=SU(2)$ for the second symbol.
The length squared of the row of the generator matrix is not $2^2+1^2$ but 
 \beqa
 \left(
\begin{array}{cc}
2~&1
\end{array}
\right)
\left(
\begin{array}{cc}
\frac12~&\\
&2
\end{array}
\right)
 \left(
\begin{array}{c}
2\\1
\end{array}
\right)
&=&4~\equiv 0~~~\mbox{mod $4$}.
 \eeqa

\subsection*{No.74 $SU(4) \times SU(4) \times SU(2) \times SU(2)$}
In this case, let us take $c_4\in\ZZ_4$, $c_2\in\ZZ_2$ and consider 
\beqa
\sum_{c_4=0}^3\sum_{c_2=0}^1
\Theta_{(c_4+c_2)\vec{\omega}^{SU(4)}_1}^{\Lambda^{SU(4)}_R}
\Theta_{(c_4-c_2)\vec{\omega}^{SU(4)}_1}^{\Lambda^{SU(4)}_R}
\Theta_{c_4\vec{\omega}^{SU(2)}_1}^{\Lambda^{SU(2)}_R}
\Theta_{c_2\vec{\omega}^{SU(2)}_1}^{\Lambda^{SU(2)}_R}(\tau).
\eeqa
The modular $S$ transformation yields
 \beqa
&=&
\frac1{(-i\tau)^{\frac{3+3+1+1}2}}
\frac
1{\sqrt{4\cdot4\cdot2\cdot2}}
\n
&&\cdot
\sum_{c_4=0}^3\sum_{c_2=0}^1
\sum_{\vec{\mu}_1\in\Lambda^{SU(4)}_W/\Lambda^{SU(4)}_R}
\sum_{\vec{\mu}_2\in\Lambda^{SU(4)}_W/\Lambda^{SU(4)}_R}
\sum_{\vec{\mu}_3\in\Lambda^{SU(2)}_W/\Lambda^{SU(2)}_R}
\sum_{\vec{\mu}_4\in\Lambda^{SU(2)}_W/\Lambda^{SU(2)}_R}
\n
&&\cdot
e^{2\pi i ((c_4+c_2)\vec{\omega}^{SU(4)}_1\cdot\vec{\mu}_1+
(c_4-c_2)\vec{\omega}^{SU(4)}_1\cdot\vec{\mu}_2+
c_4\vec{\omega}^{SU(2)}_1\cdot\vec{\mu}_3+
c_2\vec{\omega}^{SU(2)}_1\cdot\vec{\mu}_4)
}
\n
&&\cdot
\Theta_{\vec{\mu}_1}^{\Lambda^{SU(4)}_R}
\Theta_{\vec{\mu}_2}^{\Lambda^{SU(4)}_R}
\Theta_{\vec{\mu}_3}^{\Lambda^{SU(2)}_R}
\Theta_{\vec{\mu}_4}^{\Lambda^{SU(2)}_R}\left(-\frac1\tau\right).
\label{SU(4)^2SU(2)^2Theta}
\eeqa
Let
\beqa
\vec{\mu}_1=c'_4\vec{\omega}^{SU(4)}_1,~~
\vec{\mu}_2=c''_4\vec{\omega}^{SU(4)}_1,~~
\vec{\mu}_3=c'_2\vec{\omega}^{SU(2)}_1,~~
\vec{\mu}_4=c''_2\vec{\omega}^{SU(2)}_1,\eeqa
 then 
 \beqa
(\vec{\omega}^{SU(4)}_1)^2=\frac34,~~~
(\vec{\omega}^{SU(U)}_1)^2=\frac12
\eeqa
 imply
 \beqa
&&\frac
1{\sqrt{4\cdot4\cdot2\cdot2}}
\sum_{c_4=0}^3
\sum_{c_2=0}^1
e^{2\pi i ((c_4+c_2)\vec{\omega}^{SU(4)}_1\cdot\vec{\mu}_1+
(c_4-c_2)\vec{\omega}^{SU(4)}_1\cdot\vec{\mu}_2+
c_4\vec{\omega}^{SU(2)}_1\cdot\vec{\mu}_3+
c_2\vec{\omega}^{SU(2)}_1\cdot\vec{\mu}_4)
}\n
&=&\frac14
\sum_{c_4=0}^3
e^{2\pi i c_4(-\frac14 c'_4-\frac14 c''_4+\frac12 c'_2)}
\cdot
\frac12
\sum_{c_2=0}^1
e^{2\pi i c_2(-\frac14 c'_4+\frac14 c''_4+\frac12 c''_2)}.
\eeqa
The first factor is $=1\neq 0$ if
\beqa
c'_4+c''_4\equiv 2 c'_2~~~\mbox{mod $4$},
\eeqa
and then the second factor is $=1\neq 0$ if
\beqa
c'_2\equiv c''_4+c''_2~~~\mbox{mod $2$}.
\eeqa
Therefore
\beqa
\vec{\mu}_1=(-c''_4+2(c''_4+c''_2))\vec{\omega}^{SU(4)}_1,~~
\vec{\mu}_2=c''_4\vec{\omega}^{SU(4)}_1,~~
\vec{\mu}_3=(c''_4+c''_2)\vec{\omega}^{SU(2)}_1,~~
\vec{\mu}_4=c''_2\vec{\omega}^{SU(2)}_1.
\eeqa
Let us introduce $\tilde c_4:=c''_4+c''_2$, then
 \beqa
\vec{\mu}_1=(\tilde c_4+c''_2)\vec{\omega}^{SU(4)}_1,~~
\vec{\mu}_2=(\tilde c_4-c''_2)\vec{\omega}^{SU(4)}_1,~~
\vec{\mu}_3=\tilde c_4\vec{\omega}^{SU(2)}_1,~~
\vec{\mu}_4=c''_2\vec{\omega}^{SU(2)}_1.
\eeqa
Thus we find
\beqa
&&
\sum_{c_4=0}^3\sum_{c_2=0}^1
\Theta_{(c_4+c_2)\vec{\omega}^{SU(4)}_1}^{\Lambda^{SU(4)}_R}
\Theta_{(c_4-c_2)\vec{\omega}^{SU(4)}_1}^{\Lambda^{SU(4)}_R}
\Theta_{c_4\vec{\omega}^{SU(2)}_1}^{\Lambda^{SU(2)}_R}
\Theta_{c_2\vec{\omega}^{SU(2)}_1}^{\Lambda^{SU(2)}_R}(\tau)
\n
&=&
\frac1{(-i\tau)^{4}}
\sum_{\tilde c_4=0}^3\sum_{c''_2=0}^1
\Theta_{(\tilde c_4+c''_2)\vec{\omega}^{SU(4)}_1}^{\Lambda^{SU(4)}_R}
\Theta_{(\tilde c_4-c''_2)\vec{\omega}^{SU(4)}_1}^{\Lambda^{SU(4)}_R}
\Theta_{\tilde c_4\vec{\omega}^{SU(2)}_1}^{\Lambda^{SU(2)}_R}
\Theta_{c''_2\vec{\omega}^{SU(2)}_1}^{\Lambda^{SU(2)}_R}\left(-\frac1\tau\right).
\eeqa

 The coefficients of the weights of each of these theta functions 
 are obtained by multiplying 
 \beqa
\left(
\begin{array}{crcc}
1&1&~1&~0\\
1&-1&~0&~1
\end{array}
\right)
\label{tetracodegeneratormatrix}
\eeqa
by $(c_4~c_2)$ from the left.
Although we have taken $c_4\in\ZZ_4$ and $c_2\in\ZZ_2$, we can also think of $c_2$ as $\in\ZZ_4$,
then we can regard this lattice as one constructed from a code over $\ZZ_4$.
Note, however, that doing so would result in doubly overlapping lattice points.

 By coincidence, (\ref{tetracodegeneratormatrix}) coincides 
 with the generator matrix of the tetracode, although its entries are in $\ZZ_4$.

\subsection*{Case 3
 }

Finally, let us consider the case when the singularity lattice $T$ contains 
$D_{2N}=SO(4N)$ $(N=2,3,\ldots)$ in its direct-sum components. 
Nos. 64, 71, and 73 fall into this class.

\subsection*{No.73 $SO(8) \times SO(8)$}
The quotient module $\Lambda^{SO(4N)}_W/\Lambda^{SO(4N)}_R$ 
of $D_{2N}=SO(4N)$ $(N=2,3,\ldots)$ is $\ZZ_2\times\ZZ_2$.
In this case, the MW group itself is also $\ZZ_2\times \ZZ_2$ and is isomorphic to it.
However, if we take one of the direct-sum components $\ZZ_2$ as the code ring,
it is not. Still, using two $\ZZ_2$-valued symbols, we can specify an element of a 
quotient module $\Lambda^{SO(4N)}_W/\Lambda^{SO(4N)}_R$.

$SO(8)$ has rank $4$ and $\det\, C_{SO(8)}=4$. In this case, the MW group is $\ZZ_2\times \ZZ_2$.
\beqa
\sum_{c_1=0}^1\sum_{c_2=0}^1
\left(
\Theta_{c_1\vec{\omega}^{SO(8)}_3+c_2\vec{\omega}^{SO(8)}_4}^{\Lambda^{SO(8)}_R}(\tau)
\right)^2
\label{SO(8)xSO(8)Theta}
\eeqa
is modular-$S$ transformed to
\beqa
&=&
\frac1{(-i\tau)^{\frac{4+4}2}}
\frac
1{\sqrt{4\cdot 4}}
\sum_{c_1=0}^1\sum_{c_2=0}^1
\sum_{\vec{\mu}\in\Lambda^{SO(8)}_W/\Lambda^{SO(8)}_R}
\left(
e^{2\pi i (c_1\vec{\omega}^{SO(8)}_3+c_2\vec{\omega}^{SO(8)}_4
)\cdot\vec{\mu}}
\Theta_{\vec{\mu}}^{\Lambda^{SO(8)}_R}\left(-\frac1\tau\right)
\right)^2,
\eeqa
which becomes, if we let
$\vec{\mu}=c'_1\vec{\omega}^{SO(8)}_3+c'_2\vec{\omega}^{SO(8)}_4$, 
\beqa
&=&
\frac1{(-i\tau)^{4}}
\sum_{c_1=0}^1\sum_{c_2=0}^1
\frac
14
\sum_{c'_1=0}^1\sum_{c'_2=0}^1
e^{2\pi i (c_1\vec{\omega}^{SO(8)}_3+c_2\vec{\omega}^{SO(8)}_4
)\cdot (c'_1\vec{\omega}^{SO(8)}_3+c'_2\vec{\omega}^{SO(8)}_4)}
\Theta_{c'_1\vec{\omega}^{SO(8)}_3+c'_2\vec{\omega}^{SO(8)}_4}^{\Lambda^{SO(8)}_R}\left(-\frac1\tau\right)
\n
&&~~~~~~~~~~~~~~~~~~~~\cdot
\sum_{c''_1=0}^1\sum_{c''_2=0}^1
%
e^{2\pi i (c_1\vec{\omega}^{SO(8)}_3+c_2\vec{\omega}^{SO(8)}_4
)\cdot (c''_1\vec{\omega}^{SO(8)}_3+c''_2\vec{\omega}^{SO(8)}_4)}
\Theta_{c''_1\vec{\omega}^{SO(8)}_3+c''_2\vec{\omega}^{SO(8)}_4}^{\Lambda^{SO(8)}_R}\left(-\frac1\tau\right).
\n
\label{SO(8)xSO(8)ThetamodularS}
\eeqa
Since
\beqa
(\vec{\omega}^{SO(8)}_3)^2=(\vec{\omega}^{SO(8)}_4)^2=1,~~~
\vec{\omega}^{SO(8)}_3\cdot\vec{\omega}^{SO(8)}_4=\frac12,
\eeqa
we find
\beqa
&&
\frac
14\sum_{c_1=0}^1\sum_{c_2=0}^1
e^{2\pi i (c_1\vec{\omega}^{SO(8)}_3+c_2\vec{\omega}^{SO(8)}_4
)\cdot (c'_1\vec{\omega}^{SO(8)}_3+c'_2\vec{\omega}^{SO(8)}_4)}
e^{2\pi i (c_1\vec{\omega}^{SO(8)}_3+c_2\vec{\omega}^{SO(8)}_4
)\cdot (c''_1\vec{\omega}^{SO(8)}_3+c''_2\vec{\omega}^{SO(8)}_4)}
\n
&=&
\frac
14\sum_{c_1=0}^1\sum_{c_2=0}^1
e^{2\pi i(c_1(c'_1+c''_1+\frac{c'_2}2+\frac{c''_2}2)
+c_2(\frac{c'_1}2+\frac{c''_1}2+c'_2+c''_2)
)
}\n
&=&
\frac
14\sum_{c_1=0}^1\sum_{c_2=0}^1
e^{2\pi i(c_1(\frac{c'_2}2+\frac{c''_2}2)
+c_2(\frac{c'_1}2+\frac{c''_1}2)
)
}.\n
\eeqa
This is $=1\neq0$ if
\beqa
c'_1+c''_1\equiv 0~~~\mbox{mod $2$},~~~
c'_2+c''_2\equiv 0~~~\mbox{mod $2$},
\eeqa
so
\beqa
(\ref{SO(8)xSO(8)ThetamodularS})
&=&
\frac1{(-i\tau)^{4}}
\sum_{c'_1=0}^1\sum_{c'_2=0}^1
\left(
\Theta_{c'_1\vec{\omega}^{SO(8)}_3+c'_2\vec{\omega}^{SO(8)}_4}^{\Lambda^{SO(8)}_R}
\left(-\frac1\tau\right)
\right)^2,
\eeqa
which shows its modular $S$ invariance if $\frac1{\eta(\tau)^8}$ is considered together.
Modular $T$ invariance can also be verified.

As mentioned above, we define the lattice constructed by Construction A${}_{\g}$
for $\g=SO(2k)$ as 
\beqa
\Gamma_{{\cal C}_1,{\cal C}_2}&:=&\left\{
{\bf c}_1\vec{\omega}_{k-1}+{\bf c}_2\vec{\omega}_k+\vec{\bf m}
\in (\Lambda_W^{SO(2k)})^{n}
\left|\,
{\bf c}_1\in{\cal C}_1,~
{\bf c}_2\in{\cal C}_2,~\vec{\bf m}\in (\Lambda_R^{SO(2k)})^{n}
\right.
\right\}
\eeqa
for two binary codes ${\cal C}_1$, ${\cal C}_2$ of length $n$.
In the present case, the lattice can be constructed from two length-2 binary codes,
whose generator matrix can be read off as
\beqa
\left(
\begin{array}{cccc}
1&1&0&0\\
0&0&1&1
\end{array}
\right).
\eeqa

The $SO(8)$ theta in (\ref{SO(8)xSO(8)Theta}) is 
$\Theta_{0}^{\Lambda^{SO(8)}_R}(\tau)=\frac{\vartheta_3^4+\vartheta_4^4}2$ 
when $c_1=c_2=0$,
$\Theta_{\vec{\omega}^{SO(8)}_3+\vec{\omega}^{SO(8)}_4}^{\Lambda^{SO(8)}_R}(\tau)
=\frac{\vartheta_3^4-\vartheta_4^4}2$ when $c_1=c_2=1$,
$\Theta_{\vec{\omega}^{SO(8)}_3}^{\Lambda^{SO(8)}_R}(\tau)
=\frac{\vartheta_2^4}2$ when $c_1=1$ and $c_2=0$, and 
 $\Theta_{\vec{\omega}^{SO(8)}_4}^{\Lambda^{SO(8)}_R}(\tau)
=\frac{\vartheta_2^4}2$ when $c_1=0$ and $c_2=1$.
Therefore,
\beqa
(\ref{SO(8)xSO(8)Theta})&=&\left(\frac{\vartheta_3^4+\vartheta_4^4}2
\right)^2
+
\left(\frac{\vartheta_3^4-\vartheta_4^4}2
\right)^2
+
\left(\frac{\vartheta_2^4}2
\right)^2
+
\left(\frac{\vartheta_2^4}2
\right)^2\n
&=&\frac{\vartheta_3^8+\vartheta_4^8+\vartheta_2^8}2.
\eeqa

\subsection*{No.71 $SO(12)\times SU(2)\times SU(2)$}
Since
$\Lambda^{SO(12)}_W/\Lambda^{SO(12)}_R=\ZZ_2\times \ZZ_2$,
we consider
 \beqa
\sum_{c_1=0}^1\sum_{c_2=0}^1
\Theta_{c_1\vec{\omega}^{SO(12)}_5+c_2\vec{\omega}^{SO(12)}_6}
^{\Lambda^{SO(12)}_R}
(\tau)
\Theta_{c_1\vec{\omega}^{SU(2)}_1}^{\Lambda^{SU(2)}_R}(\tau)
\Theta_{c_2\vec{\omega}^{SU(2)}_1}^{\Lambda^{SU(2)}_R}(\tau).
\eeqa
The modular $S$ transformation yields
 \beqa
 &=&
\frac1{(-i\tau)^{\frac{6+1+1}2}}
\frac
1{\sqrt{4\cdot 2 \cdot 2}}
\sum_{c_1=0}^1\sum_{c_2=0}^1
\sum_{\vec{\mu}\in\Lambda^{SO(12)}_W/\Lambda^{SO(12)}_R}
\sum_{\vec{\mu}_1\in\Lambda^{SU(2)}_W/\Lambda^{SU(2)}_R}
\sum_{\vec{\mu}_2\in\Lambda^{SU(2)}_W/\Lambda^{SU(2)}_R}\n
&&\cdot
e^{2\pi i ((c_1\vec{\omega}_5^{SO(12)}+c_2\vec{\omega}_6^{SO(12)})\cdot\vec{\mu}+
c_1\vec{\omega}_1^{SU(2)}\cdot\vec{\mu}_1+
c_2\vec{\omega}_1^{SU(2)}\cdot\vec{\mu}_2
)}\n
&&
\cdot\Theta_{\vec{\mu}}^{\Lambda^{SO(12)}_R}\!\!\left(-\frac1\tau\right)
\Theta_{\vec{\mu}_1}^{\Lambda^{SU(2)}_R}\!\!\left(-\frac1\tau\right)
\Theta_{\vec{\mu}_2}^{\Lambda^{SU(2)}_R}\!\!\left(-\frac1\tau\right).
\label{SO(12)xSU(2)xSU(2)Theta}
\eeqa
If we set
\beqa
\vec{\mu}=c'_5\,\vec{\omega}^{SO(12)}_5+c'_6\,\vec{\omega}^{SO(12)}_6,~~~
\vec{\mu}_1=c'_1\,\vec{\omega}^{SU(2)}_1,~~~
\vec{\mu}_2=c'_2\,\vec{\omega}^{SU(2)}_1
~~~
(c'_5,c'_6,c'_1,c'_2\in\ZZ_2),
\eeqa
then using
\beqa
(\vec{\omega}^{SO(12)}_5)^2=
(\vec{\omega}^{SO(12)}_6)^2=\frac32, ~~~
\vec{\omega}^{SO(12)}_5\cdot\vec{\omega}^{SO(12)}_6=1,~~~
(\vec{\omega}^{SU(2)}_1)^2=\frac12,
\eeqa
we find
\beqa
&&\frac
1{\sqrt{4\cdot 2 \cdot 2}}
\sum_{c_1=0}^1\sum_{c_2=0}^1
e^{2\pi i ((c_1\vec{\omega}_5^{SO(12)}+c_2\vec{\omega}_6^{SO(12)})\cdot\vec{\mu}+
c_1\vec{\omega}_1^{SU(2)}\cdot\vec{\mu}_1+
c_2\vec{\omega}_1^{SU(2)}\cdot\vec{\mu}_2
)}\n
&=&
\frac
12
\sum_{c_1=0}^1 
e^{2\pi i c_1(\frac32 c'_5+c'_6+\frac12 c'_1)}
\sum_{c_2=0}^1
e^{2\pi i c_2( c'_5+\frac32 c'_6+\frac12 c'_2)}.
\eeqa
This is $=1 \neq 0$ if
\beqa
c'_5\equiv c'_1~~~\mbox{mod $2$},~~~
c'_6\equiv c'_2~~~\mbox{mod $2$},
\eeqa
so that
 \beqa
 &&
\sum_{c_1=0}^1\sum_{c_2=0}^1
\Theta_{c_1\vec{\omega}^{SO(12)}_5+c_2\vec{\omega}^{SO(12)}_6}
^{\Lambda^{SO(12)}_R}
(\tau)
\Theta_{c_1\vec{\omega}^{SU(2)}_1}^{\Lambda^{SU(2)}_R}(\tau)
\Theta_{c_2\vec{\omega}^{SU(2)}_1}^{\Lambda^{SU(2)}_R}(\tau)\n
&=&
\frac1{(-i\tau)^{4}}
\sum_{c'_1=0}^1\sum_{c'_2=0}^1
\Theta_{c'_1\vec{\omega}^{SO(12)}_5+c'_2\vec{\omega}^{SO(12)}_6}
^{\Lambda^{SO(12)}_R}
\!\!\left(-\frac1\tau\right)
\Theta_{c'_1\vec{\omega}^{SU(2)}_1}^{\Lambda^{SU(2)}_R}\!\!\left(-\frac1\tau\right)
\Theta_{c'_2\vec{\omega}^{SU(2)}_1}^{\Lambda^{SU(2)}_R}\!\!\left(-\frac1\tau\right).
\eeqa
This is the summation over the lattice constructed from a code over $\ZZ_2$ generated by 
\beqa
\left(
\begin{array}{cccc}
1~&0~&1~&0\\
0~&1~&0~&1
\end{array}
\right)
\eeqa
by Construction A${}_{\gsmall}$,
where $\g=SO(12)$ for the first and second symbols and 
$\g=SU(2)$ for the third and forth symbols. 

\subsection*{No.64 $SO(16)$}
$SO(16)$ has rank $8$ and $\det\, C_{SO(16)}=4$.
We consider 
 \beqa
\sum_{c=0}^1
\Theta_{c\,\vec{\omega}^{SO(16)}_7}
^{\Lambda^{SO(16)}_R}
(\tau).
\eeqa
In this case, $\Lambda^{SO(16)}_W/\Lambda^{SO(16)}_R=\ZZ_2\times\ZZ_2$, 
but only $\vec{\omega}_7$ is used to generate one $\ZZ_2$.
Its modular $S$ transform reads
\beqa
 &=&
\frac1{(-i\tau)^{\frac{8}2}}
\frac
1{\sqrt{4}}
\sum_{c=0}^1
\sum_{\vec{\mu}\in\Lambda^{SO(16)}_W/\Lambda^{SO(16)}_R}
e^{2\pi i c\,\vec{\omega}_7^{SO(16)}\cdot\vec{\mu}}
\Theta_{\vec{\mu}}^{\Lambda^{SO(16)}_R}\!\!\left(-\frac1\tau\right).
\label{SO(16)Theta}
\eeqa
If we let $\vec{\mu}=c'\,\vec{\omega}_7^{SO(16)}$, then, since $(\vec{\omega}_7^{SO(16)})^2=1$,
we have
\beqa
 &=&
\frac1{(-i\tau)^{\frac{8}2}}
\frac
1{\sqrt{4}}
\sum_{c=0}^1
\sum_{c'=0}^1
e^{2\pi i c\,c'}
\Theta_{c'\vec{\omega}_7^{SO(16)}}^{\Lambda^{SO(16)}_R}\!\!\left(-\frac1\tau\right)\n
&=&
\frac1{(-i\tau)^{4}}
\sum_{c'=0}^1
\Theta_{c'\vec{\omega}_7^{SO(16)}}^{\Lambda^{SO(16)}_R}\!\!\left(-\frac1\tau\right).
\eeqa
The generator matrix is
\beqa
\left(
\begin{array}{c}
1
\end{array}
\right).
\eeqa

\section{Conclusions}
In this paper, we discussed the relationship between error-correcting codes 
that construct the $E_8$ lattice and the Mordell-Weil group of extremal rational elliptic surfaces.
For extremal rational elliptic surfaces, the MW group becomes a torsion subgroup,
and the code defined over it encodes information about the orthogonal decomposition of $E_8$.
To construct a lattice from the code, we used a lattice construction called Construction A${}_{\gsmall}$ 
and its extensions. We have thus identified the codes that constitute the $E_8$ lattice for 
all MW lattices of extremal rational elliptic surfaces classified by Oguiso-Shioda.
The results are summarized in table II \footnote{
The kernel/cokernel columns for Nos. 64, 71, and 73 in Case 3 
indicate whether a kernel or a cokernel exists in the map
from $E(K)$ to the quotient module 
$\Lambda^{\gsmall}_W/\Lambda^{\gsmall}_R$, rather than from the code ring.}.

\setlength{\tabcolsep}{10pt}
\renewcommand{\arraystretch}{0.6}
\setlength{\arraycolsep}{3pt}
\begin{table}[h]
\caption{ \label{table2}}
\centering
\small
\begin{tabular}{|c|c|c|c|c|c|c|}
\hline
No. &$T$ &$E(K)$ & $\begin{array}{c}
\mbox{code}\\\mbox{ring}\end{array}$&
$\Lambda^{\gsmall}_W/\Lambda^{\gsmall}_R$
&$\begin{array}{c}
\mbox{kernel/}\\\mbox{cokernel}\end{array}$
& $\begin{array}{c}
\mbox{generator}\\\mbox{matrix}\end{array}$\\
\hline
$63$ &$A_8$ &$\ZZ_3$&$\ZZ_3$&$\ZZ_9$&\mbox{cokernel}&$(1)$\\
$64$ &$D_8$ &$\ZZ_2$&$\ZZ_2$&$\ZZ_2\oplus\ZZ_2$&\mbox{cokernel}&$(1)$\\
$65$ &$E_7\oplus A_1$ &$\ZZ_2$&$\ZZ_2$&$\ZZ_2$&\mbox{isomorphic}&$\left(
\begin{array}{cc}
1&1
\end{array}
\right)$\\
$66$ &$A_5\oplus A_2 \oplus A_1$ &$\ZZ_6$&$\ZZ_6$&$\ZZ_6$,$\ZZ_3$,$\ZZ_2$&\mbox{kernel}&$\left(
\begin{array}{ccc}
1&1&1
\end{array}
\right)$\\
$67$ &$A_4\oplus A_4$ &$\ZZ_5$&$\ZZ_5$&$\ZZ_5$&\mbox{isomorphic}&$\left(
\begin{array}{cc}
1&2
\end{array}
\right)$\\
$68$ &$A_2\oplus A_2\oplus A_2\oplus A_2$ &$\ZZ_3 \oplus \ZZ_3$
&$\ZZ_3$&$\ZZ_3$&\mbox{isomorphic}&$\left(
\begin{array}{cccc}
1~&~0~&~1&~~1\\
0~&~1~&~1&-1
\end{array}
\right)$\\
$69$ &$E_6\oplus A_2$ &$\ZZ_3$&$\ZZ_3$&$\ZZ_3$&\mbox{isomorphic}&$\left(
\begin{array}{cc}
1&1
\end{array}
\right)$\\
$70$ &$A_7\oplus A_1$ &$\ZZ_4$&$\ZZ_4$&$\ZZ_8$,$\ZZ_2$&\mbox{both}&$\left(
\begin{array}{cc}
2&1
\end{array}
\right)$\\
$71$ &$D_6\oplus A_1\oplus A_1$ &$\ZZ_2 \oplus \ZZ_2$
&$\ZZ_2$&$\ZZ_2\oplus\ZZ_2$,$\ZZ_2$&\mbox{isomorphic}&$\left(
\begin{array}{cccc}
1~&0~&1~&0\\
0~&1~&0~&1
\end{array}
\right)$\\
$72$ &$D_5\oplus A_3$ &$\ZZ_4$&$\ZZ_4$&$\ZZ_4$&\mbox{isomorphic}&$\left(
\begin{array}{cc}
1&1
\end{array}
\right)$\\
$73$ &$D_4\oplus D_4$ &$\ZZ_2 \oplus \ZZ_2$
&$\ZZ_2$&$\ZZ_2 \oplus \ZZ_2$&\mbox{isomorphic}&$\left(
\begin{array}{cccc}
1&1&0&0\\
0&0&1&1
\end{array}
\right)$\\
$74$ &$A_3\oplus A_1\oplus A_3\oplus A_1$ &$\ZZ_4 \oplus \ZZ_2$
&$\ZZ_4$&$\ZZ_4$,$\ZZ_2$ &\mbox{kernel}&$\left(
\begin{array}{crcc}
1&1&~1&~0\\
1&-1&~0&~1
\end{array}
\right)$\\
\hline
\end{tabular}
\end{table}

It is not clear what role the relationship between classical error codes 
and the Mordell-Weil lattices of rational elliptic surfaces considered in this paper 
will play in modern quantum information technology.
At least the Mordell-Weil lattice is a module of sections of rational elliptic surfaces, 
so it can be seen as a geometric realization of each qudit of the code, 
and may also be related (albeit in higher dimensions) to topological codes such as toric codes.

\vskip 5mm
\section*{Acknowledgments}
The work of S.M. was supported by JSPS KAKENHI Grant Number JP23K03401,
and the work of T.O. was supported by JST SPRING, Grant Number JPMJSP2104.

\end{document}